\documentclass[aps,pre,floatfix,nofootinbib,showpacs]{revtex4}
\usepackage{graphicx}
\usepackage{amsmath}
\usepackage{amssymb}

\begin{document}
\baselineskip 18pt 
\author{George Thomas and Ramandeep S. Johal\footnote{electronic address:
rsjohal@iisermohali.ac.in}}
\affiliation{Indian Institute of Science Education and Research Mohali,\\
Sector 81, S.A.S. Nagar, P.O. Manauli, Mohali 140306, India}
\draft
\title{Expected Behavior of Quantum
Thermodynamic Machines \\ with Prior Information}
\begin{abstract}
We estimate the expected behavior of a quantum model of heat engine
when we have incomplete information about external macroscopic 
parameters, like magnetic
field controlling the intrinsic energy scales of the working medium.
We explicitly derive the prior probability distribution for these
unknown parameters, $a_i, (i=1,2)$. Based on a few simple assumptions, the
prior is found to be of the form $\Pi(a_i) \propto 1/a_i$.
By calculating the expected values of various physical quantities
related to this engine, we find that the expected behavior of
the quantum model  exhibits thermodynamic-like features. 
This leads us to a surprising proposal  
that incomplete information quantified
as appropriate prior distribution can lead us to expect classical
thermodynamic behavior in quantum models.
\end{abstract}
\maketitle
\section{Introduction}
In a recent paper \cite{Johal2010}, a connection was observed  
between the notion of
prior probabilities \cite{Jeffreys1939, Jaynes1968} and the optimal performance
of certain models of quantum 
heat engines \cite{Kieu2004}. As a central feature of Bayesian inference, 
these probabilities  quantify
the uncertainty or equivalently, the lack of complete knowledge 
about some parameters of the problem. 
 It was found \cite{Johal2010} that a power-law choice of the prior,
 yields the expected optimal performance of the quantum engine
at certain well-known thermal  
efficiencies such as Curzon-Ahlborn (CA) efficiency \cite{CA1975}.
This observation is remarkable from at least two viewpoints:
i) the CA efficiency is well-known to be a finite-time thermodynamic
result, where it arises within endoreversible thermodynamic models 
as efficiency at maximum power. But the cycle considered in 
Ref. \cite{Johal2010} runs infinitely slow;
ii) the derivation of this result based
on Bayesian approach suggests a very different origin 
of these efficiencies
by connecting them with the notion of incomplete information.    

Let us clarify about the source of the assumed incomplete information.  
Usually, incomplete information about the state
of a quantum system is meant to refer to an uncertain
preparation procedure, the state being defined by a mixed 
state density matrix $\rho = \sum_i p_i | \psi_i \rangle
\langle \psi_i |$. This ensemble is characterised by
a hamiltonian $H(a)$ of the system,   
a function of externally controllable parameters 
(collectively denoted by $a$), such as magnetic field, coupling
strength with environment/subsystems and so on.
One may say that the state or the ensemble is characterised by the choice
of these external parameters. In the present discussion,
we assume an ignorance about the exact values of these
parameters. Our focus is on the following issue: 
can someone who is ignorant of the exact values of certain 
control parameters for the system, make some reasonable
estimates about its behavior? The particular physical problem that
we choose is the performance of heat engines. In particular, we ask
how should one quantify one's ignorance about the configuration
of a quantum heat engine and what are the estimates for 
its expected performance? 

Now we may possess some prior information  
about a parameter which is uncertain in the above sense,
 such as the possible range of values or, other inputs from 
the physics of the problem.
 Based on that we assign probabilities for 
 the likely values of this parameter. However, these
probabilities are to be interpreted in the 
sense of rational degree of belief \cite{Jeffreys1939}.
 Thus essentially we interpret our ignorance about parameters
of the system from a Bayesian perspective.
Formally, one may say that we are considering
an ensemble of subensembles, where each 
subensemble is characterised by a hamiltonian
$H(a)$, and the possible values of $a$ over different
subsensembles are described by a prior distribution.
We may refer to the ensemble so considered as
the {\it prior ensemble}, but the ensemble is 
regarded here as a theoretical construct only.
  Our aim is to
quantify the prior information about the 
uncertain parameters as a prior
distribution and to use this prior to make 
estimates about the behavior of the system. 
%
   
The purpose of this paper is to derive and 
justify the choice of the prior from more basic underlying assumptions 
and to further clarify the subjective approach. 
Our quantum model of a heat engine consists of a pair
 of weakly interacting two-level systems, which is closely analogous to
the model of  \cite{Johal2010, Kieu2004}. There are two reservoirs characterised
by temperatures $T_1$ and $T_2$. Additionally, we have
two energy scales, denoted by $a_i (i=1,2)$  refering to the level spacing 
of a two-level system. These spacings are controllable externally
e.g. via magnetic field, but are here assumed to be
uncertain. It is these energy scales for which we assign a prior.
The form of the prior we derive for parameter $a_i$, is  the $1/a_i$ law. 
We show the consequences of using such a prior 
for the expected values of physical quantities such as
efficiency, work and so on. We observe that the 
assignment of the prior is such that it leads us to {\it expect} a thermodynamic
like behavior in these quantum models. 

The paper is organised as follows. In Section II,
we present the model for heat engine and summarise its main features.
 In Section III, we assume  incomplete information
about internal energy scales of the working medium and so derive the 
prior for them based on some simple assumptions. In Section IV, we apply the
prior so
derived to estimate thermal efficiency of the engine,
final temperature, work extracted and so on. In Section V, we highlight a 
specific asymptotic limit in which the expected  behavior
becomes especially simple. Section VI, addresses a special case by including
a constraint. The final Section VII is devoted to conclusions and
future outlook.  
%
\section{Quantum model for work extraction}
\label{model}
Consider a pair of two-level systems labeled R and S, with hamiltonians
$H_R$ and $H_S$
having energy eigenvalues $(0,a_1)$ and $(0,a_2)$, respectively. The hamiltonian
of the composite system is given by $H = H_R \otimes I + I\otimes H_S$. 
The initial state  is $\rho_{\rm ini} = \rho_R \otimes \rho_S$,  
where  $\rho_R$ and $\rho_S $ are thermal states corresponding to
 temperatures $T_1$ and $T_2$ ($< T_1$), respectively. Let ($r_1, r_2$) and 
($s_1, s_2$) be the occupation probabilities of each system, where 
\begin{equation}
r_1 = \frac{1}{(1+e^{-a_1/T_1})}, \qquad s_1 = \frac{1}{(1+e^{-a_2/T_2})}, 
\label{}
\end{equation}
with $r_2 = (1-r_1)$ and $s_2 = (1-s_1)$. We have set Boltzmann's constant
$k_{\rm B} = 1$.
The initial mean energy of each system is
\begin{equation}
 E_{ ini}^{(i)}=\frac{a_i}{(1+e^{a_i/T_i})},
\label{enin}
\end{equation}
where $i=1,2$ denote system R and S respectively.
Within the approach based on quantum thermodynamics \cite{Gyftopoulos, ABN2004,
AJM2008},
 the process of maximum work extraction
is identified as a quantum unitary process on the thermally isolated composite
system.
It preserves not just the magnitude of von Neumann entropy of the composite
system,
but also all eigenvalues of its density matrix. It has been shown in these
works that for  $a_1 > a_2$, such a process minimises the
final energy if the final state is given by $\rho_{\rm fin} = \rho_S \otimes
\rho_R$.
Effectively, it means that in the final state the two systems \textit{swap} 
between themselves their initial probability distributions. The final energy of
each system
at the end of work extracting transformation is
\begin{equation}
 E_{ fin}^{(i)}=\frac{a_i}{(1+e^{a_j/T_j})}.
\label{enfi}
\end{equation}
where $i \ne j$. The average work per cycle defined as 
${W} \equiv {\rm Tr}[(\rho_{\rm ini}-\rho_{\rm fin})H] = E_{ini}-E_{fin}$, is
given by 
\begin{equation}
{ W}(a_1, a_2) = (a_1-a_2)  \left[ \frac{1}{\left( 1+e^{a_1/T_1}\right) } 
-  \frac{1}{\left( 1+e^{a_2 /T_2}\right) } \right].
\label{worke}
\end{equation}
To complete the cycle, the two systems are brought again in thermal contact 
with their respective reservoirs.
In this cycle, the heat extracted from the hot reservoir is
\begin{equation}
Q_1 = a_1  \left[ \frac{1}{\left( 1+e^{a_1/T_1}\right) } 
-  \frac{1}{\left( 1+e^{a_2 /T_2}\right) } \right].
\label{heat1}
\end{equation}
The efficiency of this engine $\eta =W/Q_1$ is
\begin{equation}
\eta = 1-\frac{a_2}{a_1}.
\label{et}
\end{equation}
Note that for $a_2=a_1, W=0, Q_1 > 0$ and $\eta =0$; for $a_2 = a_1 (T_2/T_1)$,
we have
the limiting values of $W =0$ and $Q_1 =0$ and $\eta = 1-(T_2/T_1)$.  
The operation of the machine as a  heat engine defined as
${ W}\ge0$ and $Q_1 \ge0$, is satisfied if
\begin{equation}
a_1 (T_2/T_1)  \le  {a_2} \le a_1.
\label{enginecriterion}
\end{equation} 
\section{Bayesian Approach}
Now consider a situation in which the temperatures of the reservoirs are
given a priori such that $T_1 > T_2$, but about the parameters $a_1$ and $a_2$
we only know that,
\begin{itemize}
\item $a_1$ and $a_2$ represent the same physical quantity, which is level
spacing
for system R and S respectively, and so they can only take positive real values.
\item If the set-up of R+S has to work as an engine, then criterion
in Eq. (\ref{enginecriterion}) must hold, whereby if one parameter is specified,
then it redefines the range of the other parameter.
\end{itemize}
Apart from the above conditions, we assume to have no information about $a_1$
and
$a_2$.
The question we address in the following is: What can we then infer about 
the expected behaviour 
of physical quantities for this heat engine (such as work per cycle,
efficiency and so on) ? 
We shall follow a subjective approach to probability to address this question. 
This implies that an uncertain parameter is assigned
 a prior distribution, which quantifies 
our priliminary expectation about the parameter to take a certain value.
 We denote the prior distribution function for our problem by $\Pi(a_1, a_2)$. 
The  prior should be assigned by taking into account any prior
knowledge or information we possess about the parameters.
For example, if $a_1$ is specified, then the prior distribution
for $a_2$, $\pi(a_2|a_1)$ is 
conditioned on the specified value of $a_1$, and is defined in the range 
$[a_1 \theta, a_1]$, where $\theta = T_2/T_1$, because we know the set-up 
works like an engine if we implement Eq. (\ref{enginecriterion}). 
 
The expected value of any physical quantity $X$ which may be function of $a_1$
and
$a_2$,  is defined as follows: 
\begin{equation}
\overline{X} = \int \int X\,\Pi(a_1,a_2)\,da_1 da_2.
\label{expectedvalue}
\end{equation}
\subsection{Assignment of the prior}
In Bayesian probability theory, 
the assignment of a unique prior is a central issue. It should quantify
not only the prior
knowledge about the parameter in the particular context, but should meet a
consistency 
criterion according to which different observers in possession of
equivalent information
should assign similar priors. Suppose, one knows only the range
within which the parameter takes its values, then 
intuition suggests that a uniform distribution may reflect
the state of our knowledge.
But such a choice is  not invariant under reparameterizations. 
Our aim in the following is to motivate the assignment of prior distributions
 for level spacings $a_1$ and $a_2$, 
when the physical problem at hand is a heat engine as described in Section
(\ref{model}).

It seems convenient to speak in terms of the two observers A and B, who wish to
assign 
priors for $a_1$  and $a_2$.
The assignment is based on the following assumptions:
\begin{description}
\item (a) Same functional form of the prior is assigned to $a_1$ and $a_2$
in the {\it initial state}, denoted by $\Pi(a_1)$ and $\Pi(a_2)$ respectively.
Further, we
assume that the prior can be expressed as $\Pi(a_i) \propto 
df(a_i)/da_i$, using a continuous differentiable function $f(a_i)$ ($i=1,2$). 
\item (b) The conditional prior distributions $p(a_j|a_i)$, implying
distribution
for $a_j$ given a value of $a_i$, or $p(a_i|a_j)$ for the converse case, have
the same functional form as above, and we assume that $p(a_j|a_i) \propto
df(a_j)/da_j$,
where the dependence on $a_i$ may be present in the normalisation factor.
Similarly, we assume $p(a_i|a_j) \propto df(a_i)/da_i$.
\end{description}

Assumption (a) is reasonable since both $a_1$ and $a_2$ represent the same
physical quantity, and the state of knowledge of A and B about them is
the same, which is the  fact that their values \textit{in the initial state} 
lie in a preassigned
range $[a_{\rm min}$, $a_{\rm max}]$. For simplicity and symmetry, we take
this range to be identical for $a_1$ and $a_2$. Presumably, this range depends  
on the experimental setup, and we assume similar
apparatus for controlling the level spacings of systems R and S.

Now if one of the parameters is specified to an observer, say $a_1$ to A,
 then A knows that
the machine works as an engine only if the range of $a_2$ is [$a_1\theta$,
$a_1$], where $a_1$ here represents some fixed value. 
Even so, assumption (b) states that A must assign the same functional form
to the prior for $a_2$ as it used for assigning to $a_1$. Although this does not
represent the 
general case, we assume this for simplicity and in the following analyse
the consequences of these assumptions.
 
Thus from (a), we have 
\begin{equation}
 \Pi(a_i)=\frac{1}{M} \frac{df(a_i)}{da_i},
\label{priorin}
\end{equation}
where the normalization constant is determined as
\begin{eqnarray} 
 M &=& \int^{a_{\rm max}}_{a_{\rm min}} \frac{df(a_i)}{da_i} da_i \nonumber \\
   & =& 
 f(a_{\rm max})-f(a_{\rm min}).
\end{eqnarray}
Following assumption (b), the  conditional probability 
distributions are given by
\begin{equation}
  \Pi(a_2| a_1)=\frac{1}{N_1}\frac{df(a_2)}{da_2},
\end{equation}
and
\begin{equation}
  \Pi(a_1| a_2)=\frac{1}{N_2}\frac{df(a_1)}{da_1},
\end{equation}
where 
\begin{eqnarray}
N_1 &=& \int_{a_1\theta}^{a_1}\frac{df(a_2)}{da_2} da_2 \nonumber \\
&=& f(a_1)-f(a_1\theta),
\end{eqnarray}
and 
\begin{eqnarray}
N_2 &=& \int_{a_2}^{a_2/\theta} \frac{df(a_1)}{da_1}  da_1 \nonumber \\
 &=& f(a_2/\theta)-f(a_2),
\end{eqnarray}
are the respective normalization constants.
 Now using the product law of probabilities, the joint prior  
$\Pi(a_1,a_2)$ as expressed by observer A
\begin{eqnarray}
\Pi(a_1,a_2)&=& \Pi(a_2| a_1)\cdot\Pi(a_1) \nonumber\\
&=&\frac{df(a_2)/da_2}{[f(a_1)-f(a_1\theta)]}\cdot\frac{df(a_1)/da_1}
{[f(a_{\rm max})-f(a_{\rm min})]},
\label{a2ga1}
\end{eqnarray}
or equivalently in terms of B, 
\begin{eqnarray}
\Pi(a_1,a_2)&=& \Pi(a_1| a_2)\cdot\Pi(a_2) \nonumber\\
&=&\frac{df(a_1)/da_1}
{[f(a_2/\theta)-f(a_2)]}\cdot\frac{df(a_2)/da_2}{[f(a_{\rm max})-f(a_{\rm
min})]}.
\label{a1ga2}
\end{eqnarray}
\begin{figure}[ht]
\begin{center}
\vspace{0.2cm}
\includegraphics[width=7cm]{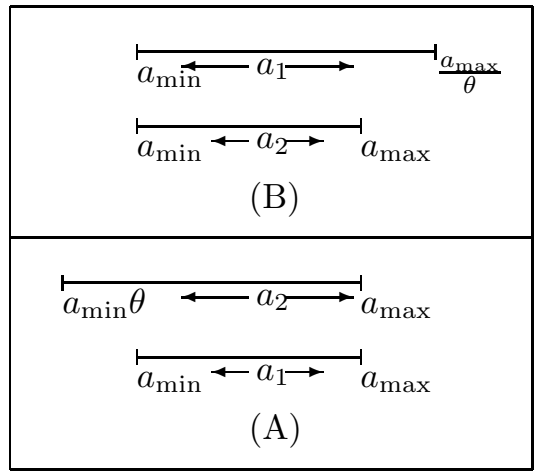}
\caption{Observer A (B) assigns the same range of values for $a_1$ ($a_2$) in the
initial state of spin R (S). But the range assigned for the other parameter 
$a_2$ ($a_1$), conditional on the operation as an engine, is different. This
fact
manifests such that A and B in general arrive at different estimates for
physical quantities.}
\hfill
\label{fig1}
\end{center}
\end{figure}
As shown in the Fig. 1, each observer assigns  different ranges of values to
$a_1$ and $a_2$.
 Each can use its own joint prior to make estimates about
any quantity. Now since either approach, that of A or B, is equivalent,
consistency
would require that they make similar estimates as to a given quantity.
Also it is reasonable to assume that for some given pair of values ($a_1,a_2$),
which
are in the allowed range of each observer, each of them should assign same 
probability for choosing $a_1$ within the small interval $da_1$ around the given
value $a_1$,
as well as of choosing $a_2$ within the small interval $da_2$, around the given
value $a_2$.
Clearly, for such a pair, both $a_1$ and $a_2$ have to lie in the interval 
$[a_{\rm min}$, $a_{\rm max}]$.
 One such pair of values,
which is definitely common to both A and B would be ($a_1,a_1$) i.e. when $a_2 =
a_1$. 
Thus equating the probabilities assigned by A and B for this case, we obtain 
from Eqs. (\ref{a2ga1}) and (\ref{a1ga2}), 
\begin{equation}
f(a_1)-f(a_1\theta)= f(a_1/\theta)-f(a_1),  
\label{f1f2}
\end{equation}
which can be rewritten as
\begin{equation}
2 f(a_1) = f(a_1\theta) + f(a_1/\theta).
\end{equation}
This functional equation has a unique solution, $f(x) = \lambda \ln x$, upto an
additive
constant. Thus we find the explicit form of prior (Eq. (\ref{priorin})) as
\begin{equation}
 \Pi(a_i) =  \frac{1}{\ln{\left(\frac{a_{\rm max}}
{a_{\rm min}}\right)}}\frac{1}{a_i},
\end{equation}
which is functionally the same as Jeffreys' choice \cite{Jeffreys1939}, also
employed in a previous study \cite{Johal2010}. 
 The joint prior is then given by
\begin{equation}
 \Pi(a_1, a_2) = 
\frac{1}{\ln\left({\frac{1}{\theta}}\right)\ln{\left(\frac{a_{\rm max}}
{a_{\rm min}}\right)}} \frac{1}{(a_1 a_2)}.
\label{jointpi}
\end{equation}
The joint prior derived above is relevant only with 
regard to the final states of R and S. As discussed in the previous section, 
the initial state of system R (S)  
depends only on parameter $a_1$ ($a_2$). This fact will be used in the following
in calculations on the expected values of quantities.

\section{Expected Values of Physical Quantities}
In this section, we use the priors assigned above, to
find expected values for various physical
quantities related to the engine. For this purpose, we employ 
the definition in Eq. (\ref{expectedvalue}). These expected 
values reflect the estimates by an observer who is  assigning the
priors. As observed in Eqs. (\ref{a2ga1}) and
(\ref{a1ga2}),
there are two methods of writing the joint prior.
So in principle, there are two ways to calculate the expected value
of some quantity and its value  will depend, in general, on the method used.
%
\subsection{Internal energy}
We calculate the expected values of internal energies for 
 systems R and S. These values can then be used to find the
expected work per cycle, heat exchanged and so on. 

(i) {\bf Initial state}: For a given $a_i$, the internal 
energy  $E_{ ini}^{(i)}$   is given by (\ref{enin}).
 The expected initial energy is defined as
\begin{equation}
\overline{E}_{ ini}^{(i)} = \int_{a_{\rm min}}^{a_{\rm max}}
 E_{ ini}^{(i)} \Pi(a_i) da_i,
\label{expectedin}
\end{equation}
where $i=1, 2$. Note that $E_{ ini}^{(i)}$ depends only on $a_i$, so we
 need to average over the prior for $a_i$ only. Thus we obtain 
\begin{eqnarray}
\overline{E}_{ ini}^{(i)}={\left[\ln{\left(\frac{a_{\rm max}}
{a_{\rm min}}\right)}\right]}^{-1}\int_{a_{\rm min}}^{a_{\rm max}}
\frac{da_i}{(1+e^{a_i/T_i})},
\label{ini}
\end{eqnarray} 
or explicitly 
\begin{equation}
 \overline{E}_{ ini}^{(i)}={\left[\ln{\left(\frac{a_{\rm max}}
{a_{\rm min}}\right)}\right]}^{-1}
\left[(a_{\rm max}-a_{\rm min})+T_i\ln{\left(\frac{1+e^{a_{\rm min}/T_i}}
{1+e^{a_{\rm max}/T_i}}\right)}\right].
\label{expein}
\end{equation}   

(ii) {\bf Final state} : In this case, the internal energy of  R as well as S,
 is function of both $a_1$ and $a_2$ (see (\ref{enfi})) and so the expected
values are
obtained by averaging over the joint prior, $\Pi(a_1,a_2)$. For instance, 
  the expected final energy of system S (denoted by superscript (2)) as
calculated by A, 
\begin{eqnarray}
\overline{E}_{fin}^{(2)}({\rm A})&=& K\int_{a_{\rm min}}^{a_{\rm max}}
\frac{1}{(1+e^{a_1/T_1})a_1}da_1\int_{a_1\theta}^{a_1}da_2,\nonumber\\
&=&K \left(1-\theta\right)
\left[(a_{\rm max}-a_{\rm min})+T_1\ln{\left(\frac{1+e^{a_{\rm min}/T_1}}
{1+e^{a_{\rm max}/T_1}}\right)}\right].
\label{EA}
\end{eqnarray}
Similarly if calculated by B,
\begin{equation}
\overline{E}_{fin}^{(2)}({\rm B})=K\int_{a_{\rm min}}^{a_{\rm max}}da_2
\int_{a_2}^{a_2/\theta}\frac{da_1}{(1+e^{a_1/T_1})a_1},
\label{efin2}
\end{equation}
where $K=[\ln{(1/\theta)}\ln{(a_{\rm max}/a_{\rm min})}]^{-1}$.
The latter integral cannot be completely solved. To simplify, 
 we rewrite it as
\begin{equation}
\overline{E}_{fin}^{(2)}({\rm B})=K\int_{a_{\rm min}}^{a_{\rm max}}da_2
\left[\int_{a_2}^{a_2/\theta}\frac{da_1}{(1+e^{a_1/T_1})a_1}\cdot 1\right].
\end{equation}
Considering the inner integral as the first function and unity as the second
function and  integrating by parts, leads to
\begin{eqnarray}
\overline{E}_{fin}^{(2)}({\rm B}) &=& K a_2\int_{a_2}^{a_2/\theta}
\frac{da_1}{(1+e^{a_1/T_1})a_1}\Bigg\vert_{a_2=a_{\rm min}}^{a_2=a_{\rm max}}
\nonumber \\
&& -  K\int_{a_{\rm min}}^{a_{\rm max}}
a_2\left[\frac{d}{da_2}\int_{a_2}^{a_2/\theta}
\frac{da_1}{(1+e^{a_1/T_1})a_1} \right]\;da_2.
\label{E2}
\end{eqnarray}
Here we use Leibniz integral rule
\begin{equation}
 \frac{d}{dy}\int_{g(y)}^{h(y)}f(x)dx=\frac{dh(y)}{dy}
f(h(y))-\frac{dg(y)}{dy}f(g(y)),
\end{equation}
to solve the second term of the Eq. (\ref{E2}) and to finally obtain
\begin{eqnarray}
 \overline{E}_{fin}^{(2)}({\rm B})&=&Ka_2 \int_{a_2}^{a_2/\theta}\frac{da_1}
{(1+e^{a_1/T_1})a_1}\Bigg\vert_{a_2=a_{\rm min}}^{a_2=a_{\rm max}}\nonumber\\
 &&-K\left[T_2\ln{\left(\frac{1+e^{a_{\rm min}/T_2}}
{1+e^{a_{\rm max}/T_2}}\right)}-T_1\ln{\left(\frac{1+e^{a_{\rm min}/T_1}}
{1+e^{a_{\rm max}/T_1}}\right)}\right].
\label{EB1}
\end{eqnarray}
In general, the expected final energies of S, as given by Eqs. (\ref{EA}) and
(\ref{EB1})  
according to  A and B, respectively, are not equal. 
One would expect that if the state of knowledge of A and B is similar,
then they should expect the same value for a given quantity. 
However, the difference in the values expected by A and B is not
so surprsing in light of the fact that different ranges for variables $a_1$ and
$a_2$
are being employed by them, as shown in Fig. 1.

Similar feature is also observed in the expressions for expected energy of
system R 
(supersript (1)), which we provide below for sake of completeness.  
\begin{equation}
 \overline{E}_{fin}^{(1)}({\rm A})=\int\int
E_{fin}^{(1)}\Pi(a_2|a_1)\Pi(a_1)\,da_2\,da_1,
\end{equation}
which implies
\begin{equation}
\overline{E}_{fin}^{(1)}({\rm A})=K\int_{a_{\rm min}}^{a_{\rm max}}da_1
\int_{a_1\theta}^{a_1}\frac{da_2}{(1+e^{a_2/T_2})a_2}.
\end{equation}
It is interesting though to observe that the above integral is {\it identical}
to the one in
Eq. (\ref{efin2}),
\begin{equation}
\overline{E}_{fin}^{(1)}({\rm A})=\overline{E}_{fin}^{(2)}({\rm B}).
\end{equation}
The second method however, yields
\begin{eqnarray} 
\overline{E}_{fin}^{(1)}({\rm B})&=& K\int_{a_{\rm min}}^{a_{\rm max}}
\frac{da_2}{(1+e^{a_2/T_2})a_2}\int_{a_2}^{a_2/\theta}da_1,\nonumber\\
&=&K\left(\frac{1}{\theta}-1\right)
\left[(a_{\rm max}-a_{\rm min})+T_2\ln{\left(\frac{1+e^{a_{\rm min}/T_2}}
{1+e^{a_{\rm max}/T_2}}\right)}\right].
\label{EA2}
\end{eqnarray}
 In the next section, we look at these expressions 
in a particular limit in which the expected values obtained by 
the two observers yield similar results, so that a meaningful 
analysis can be carried out in this limit.  
%
\section{Asymptotic Limit}
\label{asymlim}
As remarked above, observers A and B are supposed to arrive at
similar conclusions. So they should arrive at similar estimates
for physical quantities using their respective priors.
This happens in the limit, when 
 $a_{\rm min}<<T_2$ and $a_{\rm max}>>T_1$. In this limit,
 Eq. (\ref{expein}) is approximated as
 \begin{equation}
\overline{E}^{(i)}_{ini} \approx \frac{\ln 2}{\ln(\frac{a_{\rm max}}{a_{\rm
min}})}T_i.
\label{appei}
\end{equation}
The ratio $(a_{\rm max}/{a_{\rm min}})$ in the above may be large in magnitude, 
but is assumed to be finite.

Similarly, it is remarkable to note that in this limit, not only the expected
energy
of a system (R or S) calculated by either of the methods (A or B), is the same 
but also its value for system R or S is also equal. 
In particular, the first term of Eq. (\ref{EB1})
can be shown to be negligible in this limit. Thus we have (omitting 
the observer index)
\begin{equation}
\overline{E}^{(i)}_{fin} \approx
\frac{\ln 2}{\ln(\frac{a_{\rm max}}{a_{\rm min}})}\frac{(1-\theta)T_1}{\ln
(\frac{1}{\theta})},
\label{appef}
\end{equation} 
%
where $i=1,2$. Further insight into this may be obtained if we 
 estimate the final temperatures of systems R and S after the  work
extraction process. 
Now if values of both $a_1$ and $a_2$ are specified, the temperatures ($T_i'$)
of 
the two systems after work extraction, are given by \cite{AJM2008}
\begin{equation}
 T_1' = T_2 \frac{a_1}{a_2},\quad \mbox{and} \quad
T_2' = T_1 \frac{a_2}{a_1}.
\end{equation}
In general, the two final
temperatures are different from each other.
Within the present framework, when
we look at the expected values of the final temperatures as calculated by
A or B, we find
\begin{equation}
 \overline{T}_1' = \overline{T}_2' = T_1 \frac{(1-\theta)}{\ln(1/\theta)}.
\end{equation}
It is interesting to find that the assignment of the prior is such
that the two systems are expected to finally arrive at a common temperature. 
Going back to Eqs. (\ref{appei}) and (\ref{appef}) for the energies, 
we see that they satisfy a simple
relation $\overline{E}_{ini}^{(i)} \propto T_i$
and $\overline{E}_{fin}^{(i)} \propto \overline{T}_i'$.
This is reminiscient of the thermodynamic behavior of a classical ideal gas.

Next, the heat exchanged between  system $i$  and the corresponding reservoir
is given by
 $\overline{Q}_i=\overline{E}_{ini}^{(i)}-\overline{E}_{fin}^{(i)}$. 
 $\overline{Q}_i>0$ ($\overline{Q}_i<0$) represents heat absorbed 
 (released) by the system. Then the expressions for the heat exchanged with the 
reservoirs in the said limit are as follows:
\begin{equation}
\overline{Q}_1 \approx \frac{\ln2}{\ln\left(\frac{a_{\rm
max}}{a_{\rm min}}\right)} \left(1+\frac{(1-\theta)}{\ln\theta} \right)T_1,
\label{q1asym}
\end{equation}
and
\begin{equation}
\overline{Q}_2 \approx \frac{\ln2}{\ln\left(\frac{a_{\rm
max}}{a_{\rm min}}\right)} \left(1+\frac{(1-\theta)}{\theta \ln\theta}
\right)T_2.
\label{q2asym}
\end{equation}
Now the  expected work per cycle is defined as: 
 $\overline{W}=\overline{Q}_1+\overline{Q}_2$.  
Thus the  efficiency may be defined as $\eta = 1+\overline{Q}_2/\overline{Q}_1$.
Explicitly, using Eqs. (\ref{q1asym}) and (\ref{q2asym}) we get
\begin{equation}
\eta = 1 + \frac{\theta \ln\theta +(1-\theta)}{\ln\theta +(1-\theta)}.
\label{effby3}
\end{equation}
This is the efficiency at which the engine is expected to operate
for a given $\theta$. The above value is function only of the ratio
 of the reservoir temperatures and
is plotted in Fig. 2. In the limit of small temperature differences,
\begin{equation}
\eta \approx \frac{(1-\theta)}{3} + \frac{(1-\theta)^2}{9} + O(1-\theta)^3.                               
\label{expneta3}
\end{equation}
\begin{figure}[ht]
\vspace{0.2cm}
\includegraphics[width=8cm]{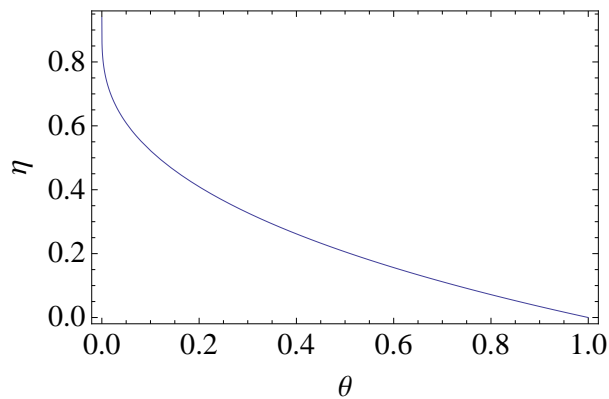}
\caption{The estimated efficiency of the engine based on expected
values of heats exchanged with the reservoirs (Eqs. (\ref{q1asym}) 
and (\ref{q2asym})). For near equilibrium,
upto first order, the efficiency is 1/3 of Carnot value.}
\hfill
\label{fig2}
\end{figure}
Before closing this section, 
 we note that the constant of proportionality in Eqs. (\ref{appei})
and (\ref{appef}),
which is ${\ln 2}\cdot\left({\ln({a_{\rm max}}/{a_{\rm min}})}\right)^{-1}$, 
can be related with heat capacity. The expected value of initial heat 
capacity of system $i$, defined as 
\begin{equation} 
\overline{C}_i = \int_{a_{\rm min}}^{a_{\rm max}}  C_i \Pi(a_i) d a_i,
\end{equation}
 where we know 
\begin{equation}
C_i=
\left(\frac{a_i}{T_i}\right)^2\frac{e^{a_i/T_i}}{(1+e^{a_i/T_i})^2}.
\end{equation}
Upon solving, the expected heat capacity  in the initial state of the system is
given exactly by
\begin{equation}
 \overline{C}_i  = \left[\ln{\left(\frac{a_{\rm max}}
{a_{\rm min}}\right)}\right]^{-1} \left[\frac{{a_{\rm max}}\,e^{a_{\rm
max}/T_i}}{T_i(1+e^{a_{\rm
max}/T_i})}
-\frac{ {a_{\rm min}}\, e^{a_{\rm min}/T_i}}{T_i(1+e^{a_{\rm
min}/T_i})} + \ln\left(\frac{1+e^{a_{\rm min}/T_i}}{1+e^{a_{\rm
max}/T_i}}\right)\right].
\label{heatcapacityini}
\end{equation}
Then in the asymptotic limit, the leading term yields  
\begin{equation}
 \overline{C}_i \equiv \overline{C} \approx\frac{\ln2}{\ln\left(\frac{a_{\rm
max}}{a_{\rm min}}\right)}.
\label{expcapacity}
\end{equation}
This limiting value is independent of temperature of the system and thus 
indicates an analogy with a constant heat capacity thermodynamic system.  

Closing this section, we note that the requirement of consistency
between the results of A and B implies, in an asymptotic limit,
that the behavior expected from minimal
prior information is the one
which shows simple thermodynamic features such as constant heat
capacity and equality of subsystem temperatures upon maximum
work extraction. 
In the next section, we revisit the above analysis, but in much
simplified form by considering an additional constraint. 
\section{Analysis at given thermal efficiency}
Let us impose the additional constraint by fixing the value of 
engine efficiency $\eta$. Then there is essentially one energy 
scale say $a_1$, to be specified 
in the model,  because the other scale $a_2$ is determined from the ratio 
$a_2/a_1 = (1-\eta)$.    
Let observer A treat $a_1$ as the uncertain parameter and denote the prior as
$\pi(a_1)$.          
The second observer B chooses $a_2$ as the uncertain parameter and the
corresponding
prior as $\pi^*(a_2)$.       
In other words, instead of $a_1$, B parametrises the uncertain quantity as 
$a_1$ times
some given constant which in the present case is $(1-\eta)$.
The probabilities assigned by A and B for a given choice of $a_1$ and
$a_2$, must satisfy
\begin{equation}
\pi(a_1) da_1 = \pi^*(a_2) da_2.
\label{pp2}
\end{equation}
On the other hand, one should expect the same functional form for the prior  
in both cases, which is saying essentially that both A
 and B are in an identical state of knowledge, implying $\pi \sim \pi^*$
\cite{Jaynes1968}.
Thus relation (\ref{pp2}) is rewritten as
\begin{equation}
\pi(a_1) = (1-\eta)\; \pi(a_1(1-\eta)),
\label{pp2p}
\end{equation}
a functional equation whose solution is given by $\pi(x) \propto 1/x$. 
Thus with the additional constraint of a given  efficiency,
the prior can be assigned as the Jeffreys' prior.  

Let us now illustrate the calculation for observer A.        
The appropriate normalised prior we have to consider is
$\pi (a_1) = \left[ { \ln \left( {a_{\rm max}}/{ a_{\rm
min}}\right)} \right]^{-1} \left({1}/{a_1} \right)$. From Eq. (\ref{worke}),
 the average work per cycle rewritten as function of $a_1$ and $\eta$ is
given by
\begin{equation}
{W}(a_1, \eta) = a_1 \eta  \left[ \frac{1}{\left( 1+e^{a_1/T_1}\right) } 
-  \frac{1}{\left( 1+e^{a_1(1-\eta) /T_2}\right) } \right].
\label{worket1}
\end{equation}
The expected work estimated by A is defined as 
$\overline{W}(A) = \int {W}(a_1, \eta) \pi(a_1) da_1$.
After calculation we have,
\begin{equation} 
 \overline{W}(A) = \left[ {\ln \left( \frac{a_{\rm max}}{ a_{\rm min}}\right)}
\right]^{-1}  \eta
\left[
\frac{T_2}{(1-\eta)} \ln \left( \frac{1+ e^{a_{\rm max}(1-\eta)/T_2}} 
     {1+ e^{a_{\rm min}(1-\eta)/T_2}}  \right )
     -T_1 \ln \left(\frac{1+ e^{a_{\rm max}/T_1}} {1+ e^{a_{\rm min}/T_1}} 
\right
)\right ].
\label{av1w}
\end{equation}
On the other hand, from the perspective of observer B who treats 
$a_2$ as the unknown parameter, the corresponding prior is
$\pi (a_2) = \left[ { \ln \left( {a_{\rm max}}/{ a_{\rm
min}}\right)} \right]^{-1} \left({1}/{a_2} \right)$. 
Upon writing the work per cycle as ${W}(a_2, \eta)$,
i.e. function of $a_2$ and $\eta$, we define 
$\overline{W}(B) = \int {W}(a_2, \eta) \pi(a_2) da_2$,
which is explicitly given by
\begin{equation}
  \overline{W}(B)=
 \left[ {\ln \left( \frac{a_{\rm max}}{ a_{\rm min}}\right)} \right]^{-1}  \eta
\left[
\frac{T_2}{(1-\eta)} \ln \left( \frac{1+ e^{a_{\rm max}/T_2}} 
 {1+ e^{a_{\rm min}/T_2}}  \right )
- T_1 \ln \left(\frac{1+ e^{a_{\rm max}/(1-\eta)T_1}}
 {1+ e^{a_{\rm min}/(1-\eta)T_1}}  \right )\right ].
\label{specialworkB}
 \end{equation}
In this case also, we see a difference in the average work expected 
by A and B, which is  at variance with our assumption that A
and B are in an equivalent state of knowledge.    
But again, we observe that in the aymptotic limit as considered in 
Section (\ref{asymlim}), both the expressions for work  get reduced to the
following simpler form
\begin{equation}
 \overline{W}(A)\approx \overline{W}(B)\approx 
\frac{\ln2}{\ln\left(\frac{a_{\rm max}}{a_{\rm min}}\right)}
\eta\left(T_1-\frac{T_2}{(1-\eta)}\right). 
\label{wspapp}
\end{equation}
It is interesting to observe that the expected work in the asymptotic limit
attains
its optimal value at the well known Curzon-Ahlborn efficiency, $\eta =
1-\sqrt{T_2/T_1}$.
This value, near equilibrium is approximated as
\begin{equation}
\eta \approx \frac{(1-\theta)}{2} + \frac{(1-\theta)^2}{8} + O(1-\theta)^3.                               
\label{expneta2}
\end{equation}
Note the difference in the above dependence on
$\theta$, with the one estimated by
Eq. (\ref{expneta3}) for the case of ignorance about two 
parameters.

Finally, we may compare the amount of work expected in the general case (as
calculated from Eqs. (\ref{q1asym}) and (\ref{q2asym}))
 when both $a_1$ and $a_2$ are uncertain, with the special case 
when the efficiency is fixed a priori. In the general case, 
the expected efficiency of the engine is given by
Eq. (\ref{effby3}), so it is reasonable to take this value
of efficiency in Eq. (\ref{wspapp}) while making
the above comparison.
 As shown in Fig. 3, the work in the general case is always less than the work
in the special case and their ratio approaches the value of 3/4 as the 
temperature gradient goes to zero ($\theta \to 1$).
\begin{figure}[ht]
\vspace{0.2cm}
\includegraphics[width=8cm]{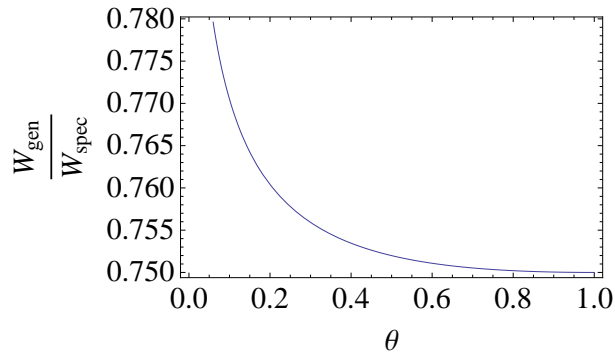}
\caption{ 
Expected work in the general case, is less than the expected work in the
special 
case for which the efficiency is fixed a priori. Both expressions for work are
calculated at the
same efficiency as given by Eq. (\ref{effby3}). As $\theta \to 1$, the ratio
approaches a
constant value 3/4.}
\hfill
\label{fig3}
\end{figure}
\section{Conclusions}
The main issue of this paper is about predictability 
in models of quantum thermodynamic machines in the presence of incomplete
knowledge about the working medium. 
The implicit reason for uncertainty considered here is different from 
other scenarios usually considered:  
The source of this uncertainty does not lie in some
unknown intrinsic dynamics of the system or due to some environment
induced fluctuations of the control parameters. We have not assumed 
any such objective mechanism which may presumably cause these values to become
uncertain. Also, the uncertainty does not stem from lack of information
about the preparation procedure for the quantum state of the working
medium. 

We have assumed a Bayesian perspective
about this (subjective) ignorance of the observer. In our model, there
are two intrinsic energy scales of the working medium, which 
are under macroscopic control, for example, via external
magnetic field. In our case, the hamiltonian itself 
is uncertain in as much as the values of control parameters
 entering in its description, are not pre-assigned.    


Once we assume a Bayesian perspective on our prior
ignorance about the system, then the task is to
narrow down the choice of prior for the uncertain parameters.
The prior so  assigned is supposed to quantify the given
prior information such as the knowledge
that these parameters are positive valued and the criterion
for operartion as a heat engine.
We make use of a consistency argument which says
that different observers in possession of similar
knowledge must assign similar priors. Invoking
two observers A and B, an appropriate
prior which we arrive at is the $1/x$ prior.  Let us emphasize that
in case further constraints
or information from data become available, then
the prior should be updated to include that 
additional information, by using suitable procedures \cite{Jaynes1968}
such as maximum entropy principle, or Bayes theorem.
In the absence of data from observations, in our opinion, the 
initial prior so assigned has to be used to make inferences
about physical quantities. The estimates for these quantities
have been defined as the average value over the chosen prior. 

Further considerations lead us to investigate a
particular asymptotic limit, because 
to maintain consistency, 
the observers A and B should arrive at similar
estimates for a given quantity, if each is 
in an equivalent state of knowledge. It is in this
limit, we observe classical thermodynamic features
for the estimated quantities of our quantum heat engine.  In particular,
the expected mean energies of the two-level
systems become proportional to their temperature, with   
  the expected heat capacity $\overline{C}$ becoming independent
 of the temperature. It is also interesting to observe that the factor 
${\ln({a_{\rm max}}/{a_{\rm min}})}$ occuring
in the normalisation of the prior can be expressed in terms of the expected
heat capacity of the system. The estimated efficiency 
behaves like one-third of Carnot value, for close to
equilibrium (nearly equal bath temperatures). 

Further, if an additional constraint
in the form of given value of efficiency is imposed, we observe that
the number of uncertain parameters in the problem reduces from
two to one.
It is then found that the expected work per cycle becomes optimal at CA
efficiency. 
In Ref. \cite{AJM2008}, it was found that CA value is a lower
bound for the efficiency at maximum work for this system.
Proximity to this value at the optimal performance has been
studied in many recent models and a certain universality has
been observed \cite{Tu2008,Esposito2010}. In this context, this 
predictability about the optimal behavior by using an appropriate
prior for ignorance, is remarkable.       

For the present model of heat engine, 
we have compared the estimated work per cycle in the general case
of ignorance of two parameters with the special case
of ignorance of one parameter only. Heuristically, one can 
anticipate that the expected work 
should increase when additional constraints are introduced which lead
to reduced uncertainty about the system. In particular, the ratio
of these two values of work, tends to a limiting value
of 3/4 as the two bath temperatures approach each other.

Similarly, the estimated efficiency close to equilibrium
for the general case is one-third of Carnot value, as contrasted
with the one-half Carnot value as found for the special case.
This may indicate that the general case of prior ignorance
about two parameters may belong to a different universality
class than the special case of ignorance about a single parameter,
which is bounded from below by half Carnot value \cite{Esposito2010,Tu2012}.
Further, we note that in a model of irreversible
Brownian heat engine \cite{Zhang2006},
 when the power is optimised with respect to the load and the barrier height, the
efficiency at optimal power is found to be given by
\begin{equation}
\eta^* = \frac{2(1-\theta)^2 }{3-2\theta(1+\ln\theta)-\theta^2}. 
 \label{effzh}
\end{equation}
Close to equilibrium, this efficiency has the same expansion as
given by Eq. (\ref{expneta3}), upto second order. 

Concluding, our analysis suggests an unexpected
relation between the average performance of
certain infinite time quantum models of heat engines infered by a Bayesian
analysis and the optimal performance of finite time thermodynamic
models.
In our opinion, the present study raises basic issues and open problems 
which we are addressing in future work, such as how robust 
is the predictability about efficiency at optimal work
by choosing other models of engines, and to further investigate
the influence of constraints on the expected behavior.
\section{Acknowledgement} RSJ acknowledges financial support from Department of 
Science and Technology, India under the project No. SR/S2/CMP-0047/2010(G).
%
 
%
%
\end{document}